\documentstyle[aps,floats,epsfig]{revtex}

\begin{document}

\twocolumn
\renewcommand{\topfraction}{1.0}
\twocolumn[\hsize\textwidth\columnwidth\hsize\csname
@twocolumnfalse\endcsname
\begin{flushright}
{\em NUB-3221-TH-01}
\end{flushright}
\title{Experimental Signature for Black Hole Production in Neutrino Air Showers}
\author{Luis Anchordoqui and Haim Goldberg}
\address{Department of Physics, Northeastern University, Boston, MA 02115, USA}

\maketitle

\begin{abstract}
The existence of extra degrees of freedom beyond the electroweak
scale may allow the formation of black holes in nearly horizontal
neutrino air showers. In this paper we examine the average
properties of the light descendants of these black holes. Our
analysis indicates that black hole decay gives rise to deeply
penetrating showers with an electromagnetic component which
differs substantially from that in conventional neutrino
interactions, allowing a good characterization of the phenomenon
against background. Naturally occurring black holes in horizontal
neutrino showers could be detected and studied with the Auger air
shower array. Since the expected black hole production rate at
Auger is $> 1$ event/year, this cosmic ray observatory could be
potentially powerful in probing models with  extra dimensions and
TeV-scale gravity.

\end{abstract}

\vskip2pc]

One of the most outstanding phenomena of TeV-scale quantum
gravity \cite{ADD} is the possible production of semi-classical
black holes (BHs) in particle collisions \cite{2,lhc1,lhc2}. These
BHs are expected to decay  promptly, giving rise to large
multiplicity events with large total transverse energy and a
characteristic ratio of hadronic to leptonic activity of roughly
5:1. Production rates for the Large Hadron Collider (LHC) are
found to be sizeable for a fundamental Planck scale $M_* = 1$ TeV
\cite{lhc1,lhc2}. Additionally, BHs occurring very deep in the
atmosphere (revealed as intermediate states of ultra high energy
neutrino interactions) may trigger quasi-horizontal showers that
could be detected with the Auger Observatory \cite{fs}. The goal
of this paper is to point out some salient experimental signatures
of these air showers.

We start the discussion by reviewing the relevant BH properties.
BHs are believed to be described by semiclassical general
relativity when their mass \mbox{$M_{\rm BH}\gg M_*$}. As $M_{\rm
BH}$ approaches $M_*$, string excitations can become important
and the BH properties rather complex. The ensuing discussion will
be framed in the context of flat extra dimensions, and we will
comment on the warped scenario after presenting our results.  In
what follows, we rely on simple semiclassical arguments assuming
that stringy effects are under control if $M_{\rm BH}/M_* \agt
5$. The Schwarzschild radius $R_{\rm S}$ of a (4+n) dimensional
BH is \cite{mp}
\begin{equation}
R_{\rm S} = \frac{1}{\sqrt{\pi}\,M_*} \left[\frac{M_{\rm BH}}{M_*} \,\,\,
\frac{8\,\,\Gamma(\frac{n+3}{2})}{n+2} \right]^\frac{1}{n+1}.
\end{equation}
Hence, if one envisions a head-on collision involving partons $i$
and $j$ with c.m. energy $\sqrt{\hat s} = M_{\rm BH}$ and impact
parameter less than $R_{\rm S}$, semiclassical reasoning suggests
that a BH is formed. The total cross section of the process can
be estimated from geometrical arguments \cite{lhc1,lhc2}, and is
of order
\begin{equation}
\hat\sigma_{ij\rightarrow BH}(\hat{s}) \approx  \pi R_{\rm S}^2 =
\frac{1}{M_*^2} \left[\frac{M_{\rm BH}}{M_*} \,\,
\frac{8\,\Gamma(\frac{n+3}{2})}{n+2} \right]^\frac{2}{n+1}.
\end{equation}
Before proceeding, we take note of a serious challenge to a
geometric cross section raised by Voloshin \cite{v}. The
criticism centers on the exponential suppression of transitions
involving a (few-particle) quantum state to a (many-particle)
semi-classical state. In response \cite{de}, the geometric result
was reaffirmed by arguing that it connects smoothly to the string
scattering cross section in an energy regime characterizing the
transition to black hole physics. Whichever point of view one may
find more convincing, it seems most conservative at this point to
depend on experiment (if possible) to resolve the issue.

With this in mind, the neutrino nucleon cross section reads
\cite{fs}
\begin{equation}
\sigma_{\nu N \rightarrow {\rm BH}} = \sum_i \int_{M_{\rm
BH}^{{\rm  min}^2}/s}^1 dx\,\, \hat\sigma_i(xs)\, \,
f_i(x,Q^2)\,\,\,,
\end{equation}
where $s = 2 m_N E_\nu$, $f_i(x,Q^2)$ are parton distribution
functions (PDFs), $M_{\rm BH}^{\rm min}$ is the minimum BH mass,
and the sum is carried out over all partons in the nucleon.
Following \cite{fs}, the cross section is calculated using the
CTEQ5M1 PDFs \cite{pdf} with the momentum transfer $Q$ taken to
be equal to $M_{\rm BH} = \sqrt{xs}$.

The energy released in neutrinos by supernova explosions imposes
several constraints on the fundamental Planck scale
\cite{bounds}. Namely, $M_*\agt 500-1600$ TeV, $M_* \agt 7-60$ TeV
and $M_* \agt 1$ TeV, for $n = 2,3,4$, respectively. Therefore, a
straightforward calculation shows that $\sigma_{\nu N \rightarrow
{\rm BH}} > \sigma_{\nu N}^{SM}$, if $n\geq 4$. Here,
\begin{equation}
\sigma_{\nu N}^{SM} (E_\nu) \approx 2.36 \times 10^{-32}
(E_\nu/10^{19}\,{\rm eV})^{0.363} \,\, {\rm cm^2}
\end{equation}
is the total charged current Standard Model $\nu N$ cross section
($10^{16}\, {\rm eV} \alt E_\nu \alt 10^{21}\, {\rm eV})$
\cite{sigma_sm}. For $M_{\rm BH} \approx 5$ TeV, $M_* = 1$ TeV,
and neutrino primary energies around $10^{20}$ eV, $\sigma_{\nu N
\rightarrow {\rm BH}} \agt 10^{-31} {\rm cm}^2$. Note that
although the atmosphere presents a target of thickness of about
1000 g/cm$^2$ to particles arriving vertically, the thickness
increases up to $\approx$ 36000 g/cm$^2$ to those arriving
tangentially to the earth surface (i.e., with horizontal
incidence to the ground). Consequently, the probability of BH
production is not negligible. Specifically, more than one BH event
per year (2 events/yr for $n=4$) could be detected by the ground
array of the Auger Observatory~\cite{fs}.

The BH lifetime, governed by the Hawking evaporation process, is
\cite{2}
\begin{equation}
\tau_{_{\rm BH}} \sim \frac{1}{M_*} \,
\left(\frac{M_{\rm BH}}{M_*}\right)^{\frac{3+n}{1+n}}.
\end{equation}
Thus, for $M_{\rm BH}\gg M_*,$ $\tau_{_{\rm BH}}M_{\rm BH}\gg 1,$
the BH is a well defined resonance and may be thought as an
intermediate state in the $s-$channel. Therefore, if one assumes
that the BH evaporates instantaneously at its original
temperature into its decay products, the average multiplicity is
roughly $M_{\rm BH}/(2\,T_{\rm H})$ \cite{lhc2}, or equivalently,

\begin{equation}
\left<N\right> \,\approx\, \frac{2\,\sqrt\pi}{n+1} \,\,
\left[\frac{M_{\rm BH}}{M_*}\right]^{\frac{n+2}{n+1}} \,\,
\left[\frac{8\,\Gamma(\frac{n+3}{2})}{n+2}
\right]^\frac{1}{n+1}\,, \label{multi}
\end{equation}
where $T_{\rm H}$ is the Hawking temperature.

Most of the large multiplicity of observable quanta emitted in the BH decay
is expected to come through hadronic jets produced by the quarks. The precise
nature of the fragmentation process is unknown. We shall use here the quark
$\rightarrow$ hadron fragmentation spectrum originally suggested by Hill
\cite{hill}
\begin{eqnarray}
\frac{dN_h}{dx} & \approx & 0.08\,\,\exp\left[2.6\sqrt{\ln(1/x)}\right]
\,\,(1-x)^2 \nonumber \\
 & \times & \left[x \sqrt{\ln(1/x)}\right]^{-1},
\label{c}
\end{eqnarray}
that is consistent with the so-called ``leading-log QCD''
behavior and seems to reproduce quite well the multiplicity growth
as seen in colliders experiments. Here, $x \equiv E/E_{\rm jet}$,
$E$ is the energy of any hadron in the jet, and $E_{\rm jet}$ is
the total energy in the jet. With  the infrared cutoff set to $x
= 10^{-3}$, the average multiplicity per jet is approximately 54
. The main features of the jet fragmentation process derived from
$dN_h/dx \, \approx\, (15/16) \, x^{-3/2}\, (1 - x)^2$ (which
provides a reasonable parametrization of Eq.(\ref{c}) for
$10^{-3} <x<1$) are listed in Table I. Now, assuming that the BH
decays into all Standard Model particles (with equipartition
among the particle species) and that each quark produces one
hadronic jet, we obtain the ``visible'' BH decay spectrum.

\begin{table}
\caption{Properties of jet hadronization}
\begin{center}
\begin{tabular}{ccccc}
$x_1$ & $x_2$  & $\int_{x_1}^{x_2} N_h\, dx$  & $\int_{x_1}^{x_2}
x\,N_h\,dx$ & $x_{\rm equivalent}$ \\ \hline
0.0750 & 1.0000 & 3  & 0.546 & 0.182  \\
0.0350 & 0.0750 & 3  & 0.155 & 0.052 \\
0.0100 & 0.0350 & 9  & 0.167 & 0.018 \\
0.0047 & 0.0100 & 9  & 0.062 & 0.007 \\
0.0010 & 0.0047 & 30 & 0.069 & 0.002 \\
\end{tabular}
\end{center}
\end{table}

We turn now to the analysis of the atmospheric cascade development
triggered by the BH secondaries. In order to propagate the particles in
the atmosphere we use the algorithms
of {\sc airesq} (version 2.1.1) \cite{sergio}.
The showering of each charged hadron in the spectrum is simulated
by a proton cascade
of energy $E$, whereas the shower induced by a $\pi^0$ decay \cite{pi0}
 is replaced
by a superposition of 2 photon showers of energy $E/2$. In $\sim
1/3$ of the events,  the leptonic channel generates
electromagnetic showers simulated by hard gamma rays. In the
remaining cases, the energy in this channel is not deposited in
the atmosphere, and we will exclude such events from our
consideration. The BH secondaries are injected with a primary
zenith angle of $80^\circ$ at 6.5 km above sea level (a.s.l.),
setting the observation level at 1.5 km a.s.l. All shower
particles with energies above the following thresholds were
tracked: 750 keV for gammas, 900 keV for electrons and positrons,
10 MeV for muons, 60 MeV for mesons and 120 MeV for nucleons. The
geomagnetic field was set to reproduce that prevailing upon the
Auger experiment. The results of these simulations were processed
with the help of the {\sc aires} analysis package. Secondary
particles of different types in individual showers were sorted
according to their distance $R$ to the shower axis. We extract in
a separate file all  $\mu^\pm$ and $e^\pm$.

\begin{figure}
\label{blaky1}
\begin{center}
\epsfig{file=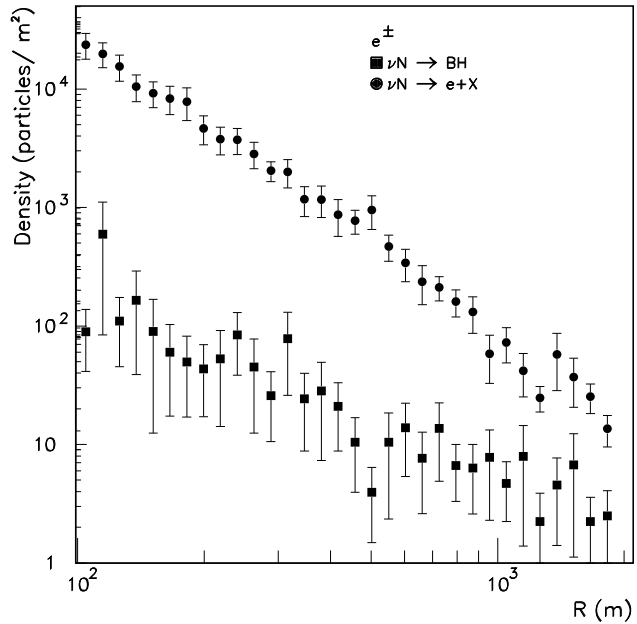,width=9.5cm,clip=} \caption{Density
distributions at ground level of $e^\pm$ as a function of the
distance to the shower axis. The primary zenith angle is
$80^\circ$ and $E_\nu = 10^{20}$ eV. The error bars indicate the
RMS fluctuations.}
\end{center}
\end{figure}

In order to obtain a clear signature of BH production, one should
be able to identify its subsequent cascade in the whole cosmic
ray sample. For large zenith angles (above $80^\circ$), an air
shower initiated by a neutrino can be distinguished from that of
an ordinary hadron by its shape. Ordinary hadrons interact high
in the atmosphere. As a consequence, at ground level the
electromagnetic part of the shower is totally extinguished (more
than 6 equivalent vertical atmosphere were gone through) and only
the muon channel survives. Besides, the shower front is extremely
flat (radius $>$ 100 km) and the particle time spread is very
narrow ($\Delta t < 50$ ns). Unlike hadrons, neutrinos may
interact deeply in the atmosphere, triggering showers in the
volume of air immediately above the detector \cite{nu}. The
shower thus presents a curved front (radius of curvature of a few
km), with particles well spread over time, ${\cal O} (\mu{\rm
s})$. If primaries are mainly electronic and muonic neutrinos (as
expected from pion decays) two types of neutrino showers can be
distinguished: ``mixed'' (with full energy) or ``pure hadronic''
(with reduced energy), respectively. In the charged current
interaction of a $\nu_e$, an ultra high energy electron is
produced which initiates a large electromagnetic cascade parallel
to the hadronic cascade.  In contrast, the charged current
interaction of a $\nu_\mu$ produces a muon which is not
detectable at Auger. For the same total energy of the primary, the
presence of a hard electromagnetic channel in BH production
provides a clean signature when compared with the ``pure
hadronic'' shower characterizing the $\nu_{\mu}$ interaction. To
analyze the differences between the BH-like shower and ordinary
$\nu_e$ shower, we mimic the latter as a superposition of a quark
jet (equivalent to the set of hadrons listed in Table I) carrying
around $20 \%$ of the original energy + a photon shower. Again,
all particles in the sample are injected at 6.5 km a.s.l. and
with a primary zenith angle of $80^\circ$.

In Fig. 1 we show the $e^\pm$ density at ground level (as a
function of their distance to the shower axis), obtained from
ordinary $\nu_e$-shower and a BH-like shower. We set $M_{\rm BH}
= 5$ TeV, and $E_\nu = 10^{20}$ eV. Then, for $4 \leq n \leq 7$,
from Eq.(\ref{multi}) we get $\left<N\right> \approx 5$. At 50 m
from the core, the ratio of the number of $e^\pm$ in a BH-like
shower to that in a typical $\nu_e$ shower is $\sim 10^{-3}$. At
about 1 km from the core this ratio rises to $\sim 10^{-1}$. Note
that the differences far from the shower-core are also
statistically significant for surface detector experiments like
the Auger Observatory \cite{prog}. In Fig. 2 we show the resulting
distributions of muons at ground level. This profile is seen to be
a rather poor discriminator between BH and ordinary showers (of
the same total energy), in spite of the fact that each of the four
hadronic jets from black hole decay has the same energy as the
single jet in the standard charged current interaction. There are
sufficient muons produced by the lepton shower to largely close
the gap between the profiles.

In the presence of maximal $\nu_{\mu}/\nu_{\tau}$-mixing,
$\nu_\tau$-showers must also be considered. However, since the
mean flight distance $\sim 50 E$ km/EeV, and the distance between
position of first impact and ground is $\sim$ 30 km, only
$\tau$'s with energy $\alt 8\times 10^{17}$ eV will decay. Thus,
$\nu_{\tau}$ showers above this energy will be indistinguishable
from $\nu_{\mu}$ showers.

We comment briefly on the warped scenario. If the curvature spills
into the extra dimensions the fundamental Planck scale can be
lowered all the way to $ \approx 1$ TeV already for $n=1.$ For BH
radii smaller than the curvature scale of the warp geometry, we
expect validity of the flat space approximation. This leads to
larger cross sections and jet multiplicities. The signature we
have described remains robust when increasing the multiplicity up
to about 15, because the properties of the electromagnetic shower
depends almost exclusively on the ratio of hadronic to leptonic
activity.

\begin{figure}
\label{blaky2}
\begin{center}
\epsfig{file=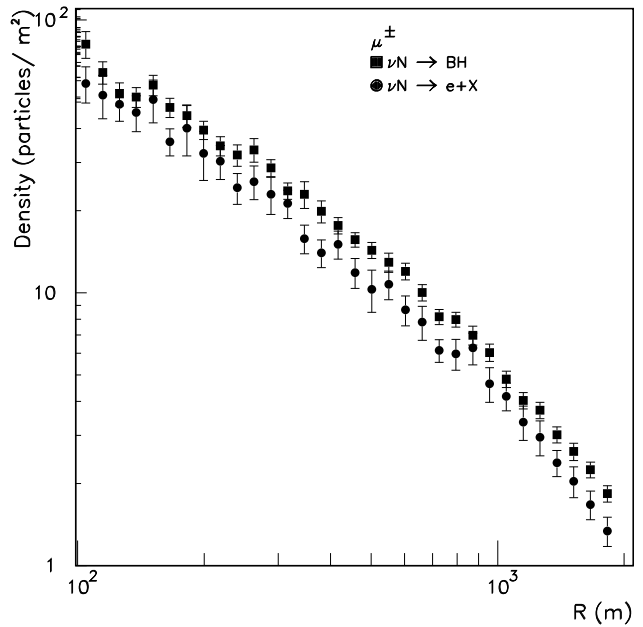,width=9.5cm,clip=} \caption{Density
distributions at ground level of $\mu^\pm$ as a function of the
distance to the shower axis. The primary zenith angle is
$80^\circ$ and $E_\nu = 10^{20}$ eV. The error bars indicate the
RMS fluctuations.}
\end{center}
\end{figure}

In summary, cosmic neutrinos with horizontal incidence to the
ground may interact with the earth atmosphere producing BHs that
decay instantaneously via Hawking evaporation. We have shown that
such a reaction chain gives rise to deeply penetrating showers
with an `anomalous' electromagnetic component: about an order of
magnitude bigger than ordinary $\nu_\mu$-showers and at least an
order of magnitude smaller than $\nu_e$-showers. This represents
a very clean signal. Our focus on a BH subsample in which the
leptonic channel generates electromagnetic showers will lower the
event rate to about 0.7/year (for $n=4,$ decreasing slowly for
larger $n).$  Thus, a 10-year collection of data at Auger could
give significant statistics to test this phenomenon, yielding
perhaps one of the early signatures of TeV-scale quantum gravity.

\acknowledgments{We would like to thank Jonathan Feng for some
valuable discussion. This work has been supported in part by the
National Science Foundation.}


\begin{thebibliography}{99}
\bibitem{pi0} Note that in our energy regime
the lifetime flight distance of any $\pi^0$ is at least an order
of magnitude smaller than the pion mean free path.
\bibitem{prog}A full simulation for projected ground level
detection at Auger is in progress.
\end{thebibliography}
\end{document}